\documentclass[aps,pra,twocolumn,showpacs,floatfix]{revtex4}
\usepackage{graphicx}

\begin{document}

\title{Entanglement of strongly interacting low-dimensional fermions in
metallic, superfluid and antiferromagnetic insulating systems}

\author{V. V. Fran\c{c}a}
\author{K. Capelle}
\email{capelle@if.sc.usp.br}
\affiliation{Departamento de F\'{\i}sica e Inform\'atica,
Instituto de F\'{\i}sica de S\~ao Carlos,
Universidade de S\~ao Paulo,
Caixa Postal 369, 13560-970 S\~ao Carlos, SP, Brazil}
\date{\today}

\begin{abstract}
We calculate the entanglement entropy of strongly correlated low-dimensional 
fermions in metallic, superfluid and antiferromagnetic insulating 
phases. The entanglement entropy reflects the degrees of freedom available
in each phase for storing and processing information, but is found not to be
a state function in the thermodynamic sense. The role of critical points,
smooth crossovers and Hilbert space restrictions in shaping the dependence
of the entanglement entropy on the system parameters is illustrated for 
metallic, insulating and superfluid systems. The dependence of the spin 
susceptibility on entanglement in antiferromagnetic insulators is obtained 
quantitatively. The opening of spin gaps in antiferromagnetic insulators
is associated with enhanced entanglement near quantum critical points.
\end{abstract}

\pacs{03.65.Ud, 03.67.Mn, 71.10.Fd, 71.10.Pm}

\maketitle

\newcommand{\be}{\begin{equation}}
\newcommand{\ee}{\end{equation}}
\newcommand{\bea}{\begin{eqnarray}}
\newcommand{\eea}{\end{eqnarray}}
\newcommand{\bi}{\bibitem}

\renewcommand{\r}{({\bf r})}
\newcommand{\rp}{({\bf r'})}

\newcommand{\ua}{\uparrow}
\newcommand{\da}{\downarrow}
\newcommand{\la}{\langle}
\newcommand{\ra}{\rangle}
\newcommand{\dg}{\dagger}

\section{Introduction}

Entanglement, one of the most surprising predictions of quantum theory,
has recently received much attention in the context of quantum information
theory and quantum computation. In quantum information theory, entanglement 
is identified with nonseparable wave functions, and referred to as "quantum 
correlation", as opposed to "classical correlation". In quantum many-body 
physics and quantum chemistry, the expression "correlation" is used in a much 
more restricted sense, meaning those consequences of the particle-particle 
interaction that arise beyond the mean-field approximation. Both meanings
of correlation are conceptually distinct: A Hartree-Fock wave function (a 
single determinant), e.g., is entangled in the quantum-information sense, 
but not correlated in the many-body sense. Even a wave function for perfectly 
noninteracting and spatially widely separated particles can be entangled (as 
illustrated by the Einstein-Podolsky-Rosen paradox), whereas correlations in 
the many-body sense vanish for strictly noninteracting particles. 

The present paper investigates correlations in the quantum information sense, 
in a model in which correlations in the many-body sense are known to be 
strong: the fermionic Hubbard model.
The one-dimensional Hubbard model in an external magnetic field has
the Hamiltonian
\bea
\hat{H}= -t\sum_{i,\sigma} (c_{i\sigma}^\dagger c_{i+1,\sigma}+H.c.)
+U\sum_i c_{i\ua}^\dagger c_{i\ua}c_{i\da}^\dagger c_{i\da}
\nonumber \\
-\mu \sum_{i\sigma} c_{i\sigma}^\dagger c_{i\sigma} 
- {h\over 2} \sum_i (c_{i\ua}^\dagger c_{i\ua} - c_{i\da}^\dagger c_{i\da}), 
\eea
where $t$ describes hopping between neighbouring sites, $U$ is the on-site
particle-particle interaction, $h$ is a magnetic field, $\mu $ the chemical
potential, and the operators $c_{i\sigma}$ satisfy fermionic commutation 
relations 
\cite{footnote1}. This Hamiltonian has a long tradition as archetypical model
of strongly interacting electrons in solids \cite{hubbardmodel,schlottmann},
comprising metallic and insulating phases at repulsive interactions, and a 
phase with superconducting correlations at attractive interactions.
More recently it has also been used to describe fermionic atoms confined in 
optical lattices \cite{atoms1a,atoms1b,atoms2}. In the absence of spatial 
inhomogeneity this model has an exact solution in terms of the Bethe-Ansatz 
(BA) \cite{hubbardmodel,schlottmann,liebwu}, which reduces
the problem of finding the many-body ground state and its energy 
to solving a set of numerically tractable coupled integral equations.

Differently from the Heisenberg model,
which is more widely used in investigations of entanglement, the Hubbard
model also accounts for the effect of itineracy of particles and possible
superconductivity. What are the effects of such complications on entanglement
measures and entanglement witnesses?

Entanglement measures for the Hubbard model were proposed and calculated, 
e.g., in Refs. \cite{gu,suecos1,suecos2,sarandy1,zanardi1,anfossi,korepin}.
Specifically, our work builds on, and extends, that of Refs. \cite{gu} and
\cite{suecos1} on the local von~Neumann entropy \cite{bennet}, or single-site 
entanglement ${\cal E}$ \cite{gu,suecos1,zanardi1} of the Hubbard model.  
Gu et al., in Ref. \cite{gu}, numerically study entanglement on
finite Hubbard clusters. For $L=70$ sites, they obtain the entanglement 
entropy as a function of $U$ precisely at haf filling ($n=1$), whereas for 
smaller clusters of up to $L=10$ sites they also obtain results for $n\neq 1$.
Larsson and Johannesson \cite{suecos1} consider the opposite limit, 
$L\to \infty$, in which they obtain the entanglement entropy of noninteracting 
particles in a half-filled band ($n=1,U=0$) and of particles with strongly 
attractive interaction in a half-filled band ($n=1,U\to-\infty$) as a 
function of magnetic field $h$, and of particles with repulsive interaction 
in zero magnetic field ($h=0,U>0$) as a function of chemical potential $\mu$
\cite{footnote2}. The main interest of Refs. \cite{gu} and \cite{suecos1}
was in the behaviour right at critical points, where the entanglement or its
derivatives were found to be strongly enhanced. 

In the present paper, we use an efficient evaluation of the Bethe-Ansatz
integral equations, recently developed in the context of density-functional
theory \cite{atoms1a,atoms1b,balda1,balda2,balda3}, in order to explore the 
entire $n(\mu)-U-h$ phase diagram, both at and away from critical points 
\cite{footnote3}. Special cases, known from the earlier work described above, 
are recovered, and complemented by new information
on the behaviour in other regions of the phase diagram. We also 
evaluate an observable not considered in Refs. \cite{gu} and \cite{suecos1}, 
namely the spin susceptibility, known to be an entanglement witness
\cite{witness1,witness2,witness3}. Our results on the spin susceptibility have 
a direct bearing on the recent reanalysis \cite{nitrate} of earlier experiments 
\cite{berger} of the entanglement in antiferromagnetic cuprate ladders, and 
also provide a new perspective on the enigmatic spin gap observed in the
normal state of cuprate superconductors.

\section{Entanglement entropy across the phase diagram}

Starting point of the present analysis is the expression of ${\cal E}$
in terms of the particle density (or filling factor) $n=n_\ua+n_\da = 
\la c_{i\ua}^\dagger c_{i\ua} \ra + \la c_{i\da}^\dagger c_{i\da} \ra $, 
the magnetization (or spin density) $m=(n_\ua-n_\da)/2$, and the
particle-particle interaction $U$,
\bea
{\cal E}\left(n,m,{\partial E\over \partial U}\right)=
-\left({n\over 2}-\frac{\partial e}{\partial U}+m\right)
\log _{2}\left[{n\over 2}-\frac{\partial e}{\partial U}+m\right]\nonumber \\
-\left({n\over 2}-\frac{\partial e}{\partial U}-m\right)
\log _{2}\left[{n\over 2}-\frac{\partial e}{\partial U}-m\right] \nonumber \\
-\left(1-n+\frac{\partial e}{\partial U}\right)
\log _{2}\left[1-n+\frac{\partial e}{\partial U}\right]
-\frac{\partial e}{\partial U}\log _{2}\left[\frac{\partial e}{\partial U}\right],
\label{tangle}
\eea
where $e=E(n,m,U)/L$ is the ground-state energy per site.

Equations equivalent to (\ref{tangle}) can be found in Refs. 
\cite{gu,suecos1}, where they are evaluated by approximating
the energy derivative $\partial E(n,m,U)/\partial U$ in certain special limits 
\cite{suecos1}, or for systems with a few lattice sites \cite{gu}, in order 
to study ${\cal E}(n,m,\partial E/\partial U)$ near critical points. Our 
interest here is in the behaviour of the entanglement entropy also away from 
critical points and special limits. To this end we must have access to the 
full function ${\cal E}(n,m,\partial E/\partial U)$. Below, we obtain this 
function from numerical solution of the BA equations for $e(n,m,U)$. In the 
context of applications of density-functional theory to spatially inhomogeneous
Hubbard models \cite{atoms1a,atoms1b,balda1,balda2,balda3} we have recently 
obtained such numerical solutions on a dense mesh of values of $n$, $m$ and $U$.

\begin{figure}
\centering
\includegraphics[height=60mm,width=70mm,angle=0]{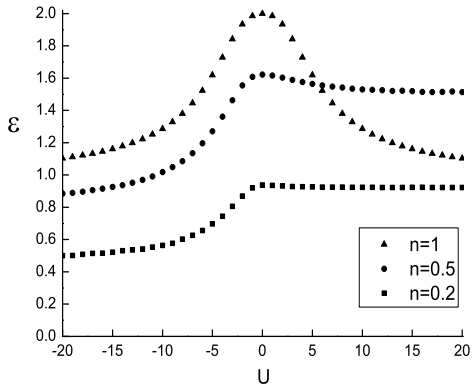}
\caption {\label{fig1}
Local entanglement entropy for $h=0$ as a function of particle-particle 
interactions, for different particle densities $n$.}
\end{figure}

\begin{figure}
\centering
\includegraphics[height=60mm,width=70mm,angle=0]{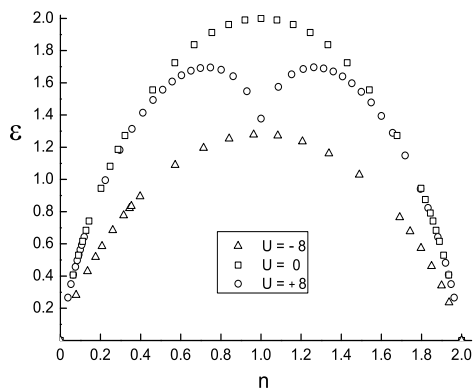}
\caption {\label{fig2}
Local entanglement entropy for $h=0$ as a function of particle density, 
for different interactions $U$.}
\end{figure}

Figures \ref{fig1} and \ref{fig2} illustrate the resulting dependence of the 
entanglement entropy on the particle-particle interaction and the particle 
density. Increase of $|U|$, which makes the wave function more correlated in 
the many-body sense, decreases ${\cal E}$, making it less entangled in the 
quantum information sense. Conversely, for noninteracting particles ($U=0$) 
the entanglement entropy reaches a maximum. This behaviour of ${\cal E}(U)$
is due to a nontrivial restriction on the Hilbert space of the one-band
Hubbard model, which permits at most one particle of each spin on each site.
This restriction remains effective even for $U=0$ (no correlations in the 
many-body sense) and implies strong correlations (entanglement) in the quantum 
information sense. We expect that similar Hilbert-space restrictions (e.g., 
due to conservation laws or topology) in more complex systems lead to other 
examples of counterintuitive relations between both notions of "correlation".

The qualitative change at the critical points $U=0$ and $n=1$ was predicted
in Refs.~\cite{gu,suecos1}, but Figs. \ref{fig1} and \ref{fig2} reveal that 
even away from critical points the dependence of ${\cal E}(n,m,U)$ displays 
considerable structure. 
For positive $U$ (repulsive interactions) the entanglement curve as a function 
of $U$ is almost flat, except at the critical point $n=1$, where a pronounced 
$U$ dependence is observed, and the curve drops from its theoretical maximum 
${\cal E}=2$ to below the values observed for $n=0.5$. This different 
behaviour at $n=1$ reflects the Mott metal-insulator transition. The rapid 
drop of the entanglement entropy in the insulating phase is due to
the freezing of electronic degrees of freedom at $n=1$ and large positive $U$,
associated with the transition from the physics of itinerant electrons to that
of localized antiferromagnetically coupled spins. At $n\neq 1$ the electrons
remain itinerant and the translational degrees of freedom are not frozen out.

We observe in Figs. \ref{fig1} and \ref{fig2} that curves corresponding
to different values of the system parameters cross, showing that physically
distinct states can give rise to the same value of the entanglement entropy,
which is thus not a faithfull state function reflecting the microscopic
structure of the ground state.

At negative $U$ (attractive interactions) a rapid drop in ${\cal E}(U)$ is 
observed for all fillings. This drop is again due to freezing of degrees of 
freedom, but at negative $U$ the relevant physics is the crossover from weakly 
coupled pairs in a BCS-like state to strongly coupled localized dimers in the 
Bose-Einstein (BE) limit. The number of degrees of freedom is reduced by a 
factor of $2$ in the formation of tightly bound pairs. Accordingly, the 
numerically determined ratio ${\cal E}(U=0)/{\cal E}(U\to-\infty) \approx 2$.
The entanglement entropy thus serves as a marker not only for 
quantum criticality \cite{gu,suecos1,sarandy1,sarandy2} but also for smoother 
crossovers. 
The strong dependence of entanglement on the size of the pairing interaction 
must be taken into account in investigations of entanglement and qubits in the 
superconducting state \cite{nakamura,mooji,zanardi2,vedral}. 

From the point of view of density-functional theory, in particular the
Hohenberg-Kohn theorem, the fundamental variables, determining the value of
any observable, are the densities $n$ and $m$, and the interaction 
$U$. From the point of view of experimental manipulation of
entangled particles in a device, it is, however, much more
convenient to cast the results in terms of variables that are more directly
controlled in the laboratory, such as the applied magnetic field $h$ or the
chemical potential $\mu$.

\begin{figure}
\centering
\includegraphics[height=60mm,width=70mm,angle=0]{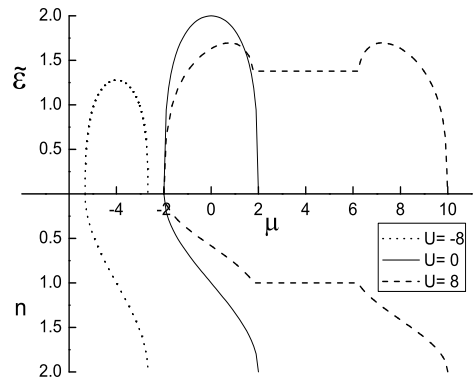}
\caption {\label{fig3}
Local entanglement entropy at $h=0$ as a function of chemical potential
(upper panel), and chemical potential as a function of density (lower panel),
for different values of the interaction $U$. Note that both panels share
a common $\mu$ axis, permiting one to reconstruct the function 
$\tilde{\cal E}(n)$.} 
\end{figure}

Given the function ${\cal E}(n,m,\partial E/\partial U)$, the relations
$\mu=\partial E(n,m,U)/\partial n$ and $h=\partial E(n,m,U)/\partial m$ can 
be used to numerically construct the dual function $\tilde{\cal E}(\mu,h,U)$.
Numerical results for $\tilde{\cal E}(\mu(n),h=0,U)$ and $\mu(n)$ are plotted 
in Fig. \ref{fig3}, which shows that 
the entanglement entropy is a very rapidly varying function of the chemical
potential. The behaviour at $U<0$ is similar to that at $U=0$, but for 
$U>0$ a gap opens at $n=1$ in the energy spectrum, the chemical potential 
becomes ill-defined, and the entanglement entropy displays a pronounced 
derivative discontinuity. 
 
\begin{figure}
\centering
\includegraphics[height=60mm,width=70mm,angle=0]{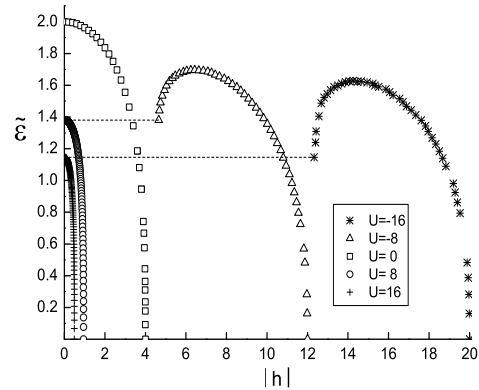}
\caption {\label{fig4}
Local entanglement entropy at $n=1$ as a function of magnetic field $h$ 
for several values of $U$.}
\end{figure}

Figure \ref{fig4} illustrates the magnetic-field dependence of the entanglement
entropy. For all $U$, the curves terminate at the magnetic field corresponding 
to magnetic saturation. For $U\ge 0$, the entanglement entropy is maximal at 
$h=0$ and decreases monotonically with increasing magnetic field, reflecting 
increased magnetic order. For $U<0$, the entropy only starts to depend on 
$|h|$ for fields above the spin gap, and displays a round maximum at some 
$|h|>0$.

\section{Spin susceptibility}

As seen above, the behaviour of the entanglement provides direct information 
about the degrees of freedom available for storing and processing 
information in systems of strongly interacting fermions. Other observables
are also strongly affected by changes in the entanglement. We here focus on
the spin susceptibility, (i) because this quantity is an example of an
entanglement witness \cite{witness1,witness2,witness3}, and (ii) because of 
recent claims \cite{nitrate} that anomalies earlier observed \cite{berger} 
in the spin susceptibility of the antiferromagnetic (AFM) insulator 
$\rm{Cu(NO_2)_22.5H_2O}$ are due to enhanced entanglement. This claim is based 
on the observation that a puzzling and hitherto unexplained \cite{berger} 
drop in the spin susceptibility begins to develop at approximately the same 
temperature where entanglement is predicted to increase \cite{nitrate}. 

However, as already indicated in the experimental paper \cite{nitrate}, many 
different explanations for the observed anomaly are conceivable. 
Moreover, in recent analysis of experiments on $\rm{LiHo_xY_{1-x}F_4}$, with
predominantly ferromagnetic interactions, precisely the opposite behaviour,
namely an increase in the spin susceptibility as entanglement becomes stronger,
was argued to occur \cite{nature}.

It is thus
important to verify the connection between the susceptibility and
entanglement in a model stripped off all unessential complications, but still
complex enough to encompass both antiferromagnetism \cite{nitrate}
and entanglement. The spins in $\rm{Cu(NO_2)_22.5H_2O}$ are 
arranged in chains, so that a description in terms of the one-dimensional 
Hubbard model, which at $n=1$ and $U\to \infty$ is equivalent to the AFM 
Heisenberg model with $J=4t^2/U$, becomes possible.

\begin{figure}
\centering
\includegraphics[height=60mm,width=70mm,angle=0]{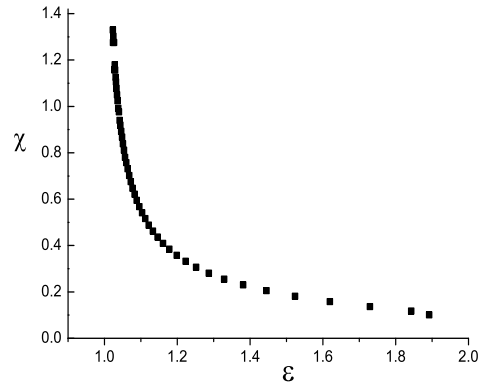}
\caption {\label{fig5}
Spin susceptibility at $n=1$ and $h=0$ as a function of the
entanglement entropy.}
\end{figure}

The zero-temperature spin susceptibility of the Hubbard model can be related 
to the ground-state energy via 
$\chi^{-1} = \partial^2 E(n,m,U)/\partial m^2|_{m=0}$, and may be obtained 
numerically from the Bethe Ansatz equations \cite{shiba} as a function of $U$. 
From Fig. \ref{fig1} we see that for $U>0$ and $n=1$ the function 
${\cal E}(U)$ is invertible. Hence, we can numerically construct the function 
$\chi^{-1}(U({\cal E}))=\chi^{-1}({\cal E})$. For $n=1$ (consistent with 
the fact that the experiments were done on insulators), this function is 
plotted in Fig. \ref{fig5}, which clearly shows that the susceptibility 
decreases with increasing entanglement entropy. Hence, in a minimal (but still
realistic) model of antiferromagnetic insulators the susceptibility indeed
drops sharply as the system's wave function becomes more entangled.
This result strongly suggests that the observation of a drop in the spin 
susceptibility in $\rm{Cu(NO_2)_22.5H_2O}$ \cite{berger} as entanglement
grows \cite{nitrate} is not a coincidence, or an artifact.

These Hubbard-model calculations, and according to \cite{nitrate} also the 
experiments of Ref. \cite{berger}, show that enhanced entanglement in AFM 
systems produces a sharp drop in the spin susceptibility. The Hubbard-model 
calculations reported in the first part of this Letter, as well as related work
in \cite{gu,suecos1,sarandy1}, show that in the proximity of quantum-critical 
points the entanglement entropy or its derivatives can be very strongly 
enhanced. By putting these two pieces of information together we infer that
in a strongly entangled state the spin susceptibility of antiferromagnetic 
systems is suppressed near quantum critical points.

Interestingly, there is another important class of materials that display
this type of behaviour, although it is not normally discussed in these terms:
The parent compounds of high-temperature superconductors are AFM cuprates,
for which much experimental evidence points to the existence of a 
quantum-critical point. The spin susceptibility of these systems displays a 
puzzling drop at low temperatures (see, e.g., Fig.~4 and p.~16004 of 
\cite{nakano}, and the recent re-analysis of those data in Fig.~4 of 
\cite{lee}). This drop is very similar to the one observed in the AFM spin 
chain $\rm{Cu(NO_2)_22.5H_2O}$, and suggests that the opening of a spin gap 
in high-temperature superconductors is also be associated with entanglement, 
persisting at relatively high temperatures. 

\section{Conclusions}

In summary, we have confirmed the previous observation \cite{gu,suecos1,suecos2,sarandy1,zanardi1} that the entanglement entropy, or its derivatives, are 
strongly enhanced near quantum critical points. 
Our Bethe-Ansatz-based treatment 
additionally reveals that (i) similar behaviour is observed at crossovers, 
(ii) even away from critical points or crossovers, each physically different
phase leaves its distinctive mark on the entanglement entropy, (iii) but in 
spite of this, the entanglement entropy is not a state function in the 
thermodynamic sense. It does, however, (iv), provide detailed information 
about which degrees of freedom are available for storing and processing 
information in each of the possible phases of the system. 
(v) Hilbert-space restrictions can lead to counterintuitive relations between 
the concept of correlation, as employed in many-body physics, and that of 
entanglement, employed in quantum-information theory.

In addition to the entropy, we also extract another observable closely
related to entanglement: the spin susceptibility. We find numerically that 
in AFM systems enhanced entanglement sharply reduces the spin 
susceptibility, in agreement with recent reanalysis \cite{nitrate} of earlier 
experiments \cite{berger} but in sharp contrast to what was found in 
ferromagnetic systems \cite{nature}. This observation suggests that the
enigmatic spin gap, observed above the critical temperature in cuprate 
superconductors, may also be a consequence of enhanced entanglement due to 
proximity of a quantum critical point. Note that high-temperature entanglement 
in cuprates was suggested on different grounds also in \cite{vedral}.

{\bf Acknowledgments}
This work was sup\-por\-ted by FAPESP, CNPq and CAPES. We thank V. L. Campo Jr. 
for help with the numerical solution of the BA equations.

\end{document}